\documentstyle[11pt,paspconf]{article}

\begin{document}
\input{psfig.tex}

\title{An Overview of Blazar Variability}

\author{C. Megan Urry}
\affil{Space Telescope Science Institute, 3700 San Martin Drive,
	Baltimore, MD 21218}

\begin{abstract}
Blazars are characterized by rapid variability at virtually all wavelengths
from radio through TeV gamma-rays. The challenge since their discovery has 
been to understand the origin of their luminous, apparently nonthermal, nuclear
emission. Considerable progress has been made in recent years thanks to a
handful of multiwavelength monitoring campaigns with high enough temporal
sampling to resolve the most rapid variations. The best data for a few objects
have shown a variety of behaviors, for the most part commensurate with
synchrotron and Compton-scattered emission from a relativistic jet, though
better data for more blazars are still clearly needed. In particular,
the origin of the seed photons that are upscattered to gamma-ray energies
remains unclear. The latest multiwavelength light curves for the BL~Lac
object PKS~2155--304 appear to rule out synchrotron emission from a
homogeneous source.
\end{abstract}


\section{Introduction}

Soon after blazars were discovered and identified, they were selected as interesting
targets for long-term monitoring programs, by pioneers like Alex
Smith among others, because of their extreme characteristics:
they were the most variable, the most luminous, the most polarized, 
and in some sense the most exciting type of Active Galactic Nuclei (AGN). 
Understanding them at first seemed the key to understanding the AGN phenomenon.
Ironically, blazars were eventually perceived as less interesting for the 
very same reason --- because they were so unusual, so the reasoning went, 
they must not be relevant to the greater body of AGN (i.e., Seyfert galaxies
and quasars). Blazar research was seen as a special field and blazars
as arcane oddities.

Now in the mid-90s, we have come full circle. Because blazars are
rare geometric manifestations of a general phenomenon (assuming
they are relativistic jets pointing directly at us; Urry \& Padovani 1995),
they must be a quite common kind of AGN, and 
when they are not pointing at us, we simply call them radio galaxies.
This makes blazars highly relevant to understanding AGN as a whole.
Specifically, the enormous energy of a relativistic jet, 
its emanation from the vicinity of the putative central black hole,
and its high degree of collimation over many orders of magnitude in
scale, offer direct clues to the extraction of energy from the
black hole. Blazars thus reveal the energetic processes occurring in 
the very centers of active galaxies, while in the more common 
radio galaxies, jet radiation (and hence information) is beamed
away from us.
With blazars, then, the goal is to understand black hole
physics through understanding the physics of the jet. This in turn
can be deduced from multiwavelength spectral characteristics, most notably
correlated variability across the spectrum.

\section{Blazars as Relativistic Jets}

Here I summarize the arguments for believing blazars
are relativistic jets. First, they commonly exhibit superluminal motion
(Vermeulen \& Cohen 1994; see also Wehrle et al., these Proceedings)
which, while it could arise from pattern rather than bulk 
relativistic velocity, is at least suggestive. Second, blazars
can exhibit extremely high brightness temperatures; in at least
some cases, intraday variations are observed at optical wavelengths,
ruling out an extrinsic (scintillation) explanation for the variability,
although the implied bulk Lorentz factors are uncomfortably high
(Wagner \& Witzel 1995).
Third, the characteristically high and variable polarization of blazars
is explained naturally by an aligned jet (Smith et al., these Proceedings).
Fourth, multiwavelength radio variability is well explained by shocks in a jet 
(Aller, these Proceedings).

Fifth and most compelling, the strong and variable gamma-ray emission 
observed in many blazars (Hartman, these Proceedings) implies such 
a high compactness that the gamma-ray source would inevitably
be dominated by pair production unless the emission is relativistically
beamed (Dondi \& Ghisellini 1995). While there is some uncertainty
about the degree of beaming required (it depends on the 
ambient X-ray photon density, as X-gamma interactions are the most
likely pair-production mechanism), the argument for some beaming is
fairly tight.

Blazars can be defined in various ways, via their rapid variability,
their compact flat-spectrum radio emission, their superluminal motion, 
their polarization, and now their gamma-ray brightnesses, 
and in fact these characteristics occur in the same sources.
That is, those sources that are superluminal have flat radio spectra 
and are highly polarized, and so on.\footnote{An exception 
is that some highly polarized quasars (HPQ), largely radio-quiet,
have continuum emission polarized by scattering rather than intrinsic
processes like synchrotron radiation. These obviously do not have 
blazar characteristics like superluminal motion or rapid variability.}
An even more direct link has been seen in at least two cases.
Ten years of VLBI maps of 3C~279 (Wehrle et al., these Proceedings) 
show a new VLBI component being ``born'' 
(extrapolating the observed position backward with the observed velocity)
at the time of the bright gamma-ray flare in June 1991 (Wehrle et al.
1994). The same phenomenon has been observed in PKS~0528+134 (Pohl et al. 1995).
In these two blazars, the production of (beamed) gamma-rays 
is directly related to superluminal motion of the radio source.
That these various blazar characteristics are all closely linked,
statistically and in some cases directly, is
a strong argument that the underlying cause is relativistic beaming.

\section{Multiwavelength Spectra and Monitoring}

\begin{figure}
\centerline{\psfig{figure=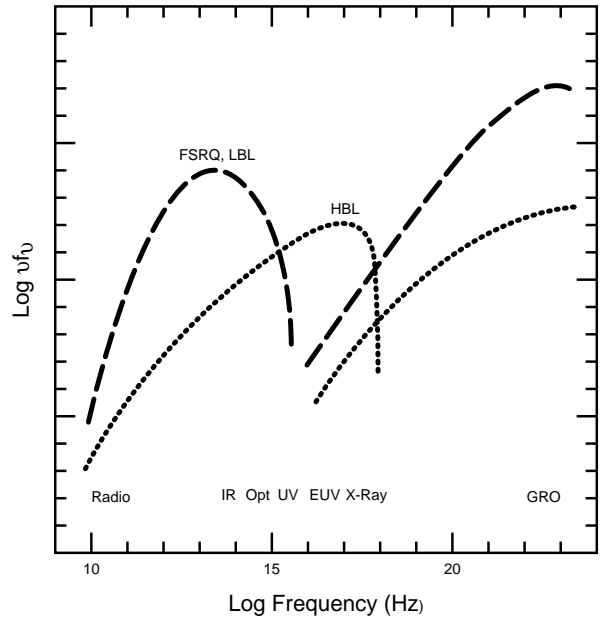,height=3.2in}}
\caption{Schematic broad-band spectra of bla\-zars from ra\-dio through
TeV gamma rays. The low-energy component
is probably due to synchrotron radiation and the high-energy component to Compton
scattering of lower-energy seed photons, possibly the synchrotron photons
or ambient UV/X-ray disk or line photons. Two different curves represent
the average spectral shapes (Sambruna et al. 1996)
of HBL (High-frequency peaked BL Lac objects; dotted line) 
and LBL (Low-frequency peaked BL Lac objects; dashed line) 
as defined by their ratios of X-ray to radio flux (see footnote 2).
Strong emission-line blazars (i.e., flat-spectrum radio quasars, or FSRQ) 
have continua like LBL (Sambruna et al. 1996).} \label{fig:bbspec}
\end{figure}

Observations in individual wavebands have established the viability
of the relativistic beaming hypothesis for blazars but have led
to at best cursory understanding of the physical state of the jet.
For this, multiwavelength variability holds the key. Blazar spectra span
an extremely broad range of energies, 
from radio through GeV gamma-rays and perhaps through TeV gamma-rays.
The emission consists of
two distinct spectral components, a low-energy synchrotron bump 
and a high-energy Compton-scattered bump (Figure~\ref{fig:bbspec}).
The temporal evolution of each spectral component and the correlation between
them are critical to understanding the underlying emission mechanisms,
which is why a number of large multiwavelength monitoring campaigns have
been carried out in recent years. 

It should be noted that the difficulty of arranging these 
mul\-ti\-wave\-length mo\-ni\-tor\-ing ob\-ser\-va\-tions, 
ty\-pi\-cal\-ly co\-or\-di\-nat\-ed among several satellites
and many more ground-based telescopes, has kept us from obtaining good
data on more than a few of the brightest blazars. More importantly,
sensitivity limits have introduced significant target selection effects.
Specifically, among BL Lac objects, 
``High-frequency peaked BL Lacs'' (HBL) and
``Low-frequency peaked BL Lacs'' (LBL) have
distinctly different continuum shapes.\footnote{The
two sub-classes of BL Lac object are defined by their
ratio of X-ray to radio flux, which anti-correlates
with the wavelength of their peak synchrotron emission.
A High-frequency peaked BL Lac (HBL) has $\alpha_{\rm rx}$
(between 5~GHz and 1~keV) less than 0.75, while a
Low-frequency peaked BL Lac (LBL) has $\alpha_{\rm rx} > 0.75$
(Padovani \& Giommi 1995).}
These may indeed be opposite extremes of a continuous distribution because
current BL Lac samples come from radio or X-ray surveys
with fairly high flux limits and so are naturally
dominated by LBL or HBL, respectively.
The strong emission-line blazars (i.e., flat-radio-spectrum quasars, or FSRQ),
also generally radio-selected, have continua like LBL (Sambruna et al. 1996).

In any case, the spec\-tral en\-er\-gy dis\-tri\-bu\-tions 
of HBL and LBL/FSRQ differ in several ways (Fig. 1).
The peak wavelength of the synchrotron component is in the 
infrared-optical band for LBL and FSRQ, 
whereas it peaks in the extreme ultraviolet to soft X-ray range for HBL. 
Also, LBL and FSRQ have a higher ratio of gamma-ray to synchrotron 
flux than the HBL (Sambruna et al. 1996), and most of the EGRET blazars
are in fact FSRQ and LBL.
Note that the synchrotron emission is most variable above the peak 
in $\nu F_\nu$, where the shortest wavelength component becomes 
optically thin (Ulrich, Maraschi, \& Urry 1996).

Only two blazars, Mrk 421 and Mrk 501, both HBL, 
have been detected at TeV energies (Punch et al. 1992, Quinn et al. 1996). 
It may be that these
particular objects were detected because they are relatively nearby
($z\sim0.03$ in both cases) so that the ultra-high-energy gamma-rays 
have little path length along which to produce pairs via scattering of 
intergalactic microwave photons, but it is probably also significant
that the peak Compton emission in HBL is likely at much higher energies
than in LBL/FSRQ. The luminosity represented by the extension of the 
blazar spectrum to GeV/TeV energies is phenomenal; clearly, 
understanding the production of this emission is central to 
understanding the blazar.

Because of these systematic spectral differences, the 
observational details of the multiwavelength
study dictate the type of blazar studied. To study blazars above the 
synchrotron peak,
where they are most variable, means selecting UV- and X-ray-bright targets,
which are inevitably HBL. To study correlated intraday variability at 
radio and optical wavelengths means selecting LBL/FSRQ. 
To correlate with GeV gamma-rays, one looks primarily at LBL/FSRQ; 
to correlate with TeV gamma-rays, one looks instead at HBL.
With higher sensitivities, this artificial distinction will disappear,
but it is important to remember that as presently observed,
the radio-optical intraday variables are systematically different 
objects than the highly variable UV/X-ray-bright sources.

With multiwavelength monitoring of blazars,
there are two critical questions we are in the process of addressing.
First, where do the gamma-rays come from, and second, what is the structure 
of the jet itself? 
By figuring out what the particle density is, 
what the magnetic fields are, how each varies along the jet, and
what causes flaring behavior, we can ultimately understand what created 
the jet, how it was formed, and what is happening down at the center 
where we cannot observe directly.
For the rest of this paper, I discuss
only two objects, 3C~279 and PKS~2155--304, which
illustrate some of the best available data (see also Takahashi et al.,
these Proceedings) and thus the limits of what we can
learn about blazars at this point.

\section{Multi-Epoch Flaring in the Superluminal Quasar 3C~279}

\begin{figure}
\centerline{\psfig{figure=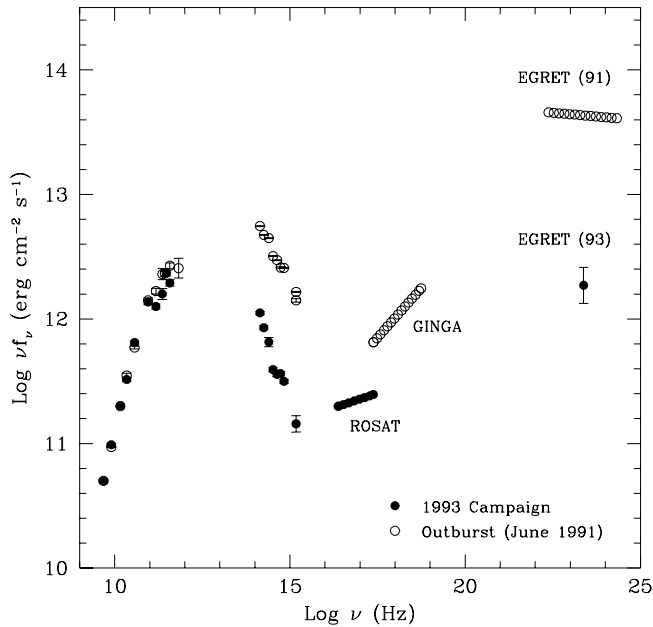,height=3.5in}}
\caption{Multi-epoch broad-band spectra of 3C~279, in a high state in
June 1991 and in a low state in January 1994
(Maraschi et al. 1994). While the UV decreased by
a factor of 3-4, the gamma rays decreased by a factor of $\sim10$ or more.
As is typical for blazars, there is little or no variability below the peak
wavelength of the synchrotron component, the peak of which is in the 
unobserved far-infrared (now accessible with ISO).}
\label{fig:3c279}
\end{figure}

3C~279 is the brightest gamma-ray blazar in the sky. [During the week
of the Miami blazar meeting, it was undergoing a major outburst, 
rising an order of magnitude above its previous highest gamma-ray state,
with substantial variations at other wavelengths.] Figure~\ref{fig:3c279} shows 
the broad-band spectra at two epochs, the high state in June 1991 when it
was first discovered and a low state in January 1994; 
the greatest change in intensity occurs at gamma-ray wavelengths, 
with lesser but still substantial variations at UV and
X-ray wavelengths, and little change at radio to sub-millimeter wavelengths
(Maraschi et al. 1994).
Note that the gamma-ray flux can dominate the bolometric luminosity, 
particularly in the flare state. Maraschi et al. (1994) suggested that
the larger increase in gamma-rays relative to synchrotron emission was
consistent with the synchrotron self-Compton (SSC) model, although a change 
in Doppler factor of the jet or concurrent increases in seed photons
and scattering electrons could also cause the observed variability.
Clearly, sampling these flares more finely is critical to identifying
the origin of the seed photons and thus the production of the dominant
spectral component. The multiwavelength observations of the 1996 flare 
should help resolve the ambiguity about the identity of the seed photons
(Wehrle et al., in preparation).

The basic blazar models currently under consideration have one common element,
a syn\-chro\-tron-emit\-ting jet filled with energetic electrons and perhaps
positrons. In addition, there may be ambient UV and X-ray emission from 
the vicinity of an accretion disk, and broad-line emission from clouds
farther out, photoionized either by the central UV/X-ray source or by
the jet itself. 
The debate turns on which photons are Compton-scattered to gamma-ray energies. 
Photons impinging on the jet from the side are boosted
in the frame of the jet electrons, and so constitute a very intense 
flux of seed photons even when the directly observed non-jet 
UV/X-ray flux is low.

Ghisellini and Madau (1996) present a nice comparison of the principal ideas
(see references therein as well) and conclude two interesting points. 
One is that the inner part of the jet has to be dissipationless.
If the energy density of gamma-rays 
in an optically thick inner region of the jet were high,
there would probably be enough local UV photons to generate a
pair cascade, transferring much of the gamma-ray energy into X-rays, contrary
to what is observed. So the principal mode by which energy is transferred
from the black hole to the jet must not be via energetic photons.
The second point is that, for plausible numbers, the illumination of the 
broad-line clouds by the beamed continuum can be an important 
contribution to the UV flux impinging on the jet. 
The bulk of the gamma-rays could therefore come from scattered broad-line 
photons which were photoionized by the jet itself. This might be the
dominant mechanism in FSRQ, say, while SSC emission dominates in 
the weaker-lined HBL.

It would be extremely interesting to monitor simultaneously the variability 
of the broad lines and the synchrotron continuum. As far as I am aware
this has not yet been done, at least with sufficient sampling. We have 
looked at the archival UV data for 3C~279 (Koratkar et al. 1996)
but even there, with a very
well-observed source, the data are not sufficient to determine the
photoionization source unambiguously. The 1996 campaign 
on 3C~279 will help and there are a number of other blazar campaigns 
planned for the upcoming year which may contribute to solving this problem.
Unfortunately, the loss of IUE, with its long, sustained monitoring
campaigns, is a major blow for this kind of study.

\section{Multiwavelength Variability of the HBL PKS~2155--304}

PKS~2155--304 is the brightest BL Lac object at ultraviolet wavelengths
and one of the brightest in the X-ray as well. So it is the obvious
choice for UV/X-ray monitoring. It is an HBL, with peak synchrotron
emission near $10^{17}$~Hz, which is four orders of magnitude higher
in frequency than the peak synchrotron emission in 3C~279.
We have every reason to expect, therefore,
that the physics of the emission from PKS~2155--304 and 3C~279 differ
in significant ways.

There have been two intensive multiwavelength campaigns to observe 
PKS 2155--304. [There was a third three months after the
Miami blazar meeting.] The first was in November 1991 and lasted for one month. 
At that time, no one even knew whether the UV and X-ray emission were related,
nor what the fastest time scale for variability was. 
We observed PKS~2155--304 with IUE once per day throughout November 1991
(thinking that was probably overkill) and then in the middle of the month, 
at the insistence of Rick Edelson (who rightly realized it was not overkill),
we observed it for nearly 5 days continuously. The UV variations 
we detected
were indeed fast enough that the daily sampling was insufficient and only
the continuous observations were useful for multiwavelength 
cross-correlations (Urry et al. 1993). 
We also had $\sim3.5$~days of continuous X-ray observations (Brinkmann et 
al. 1995) overlapping with most of the intensive UV coverage, as well as
considerable ground-based radio, infrared, and optical observations
(Smith et al. 1992, Courvoisier et al. 1995).

\begin{figure}[t]
\centerline{\psfig{figure=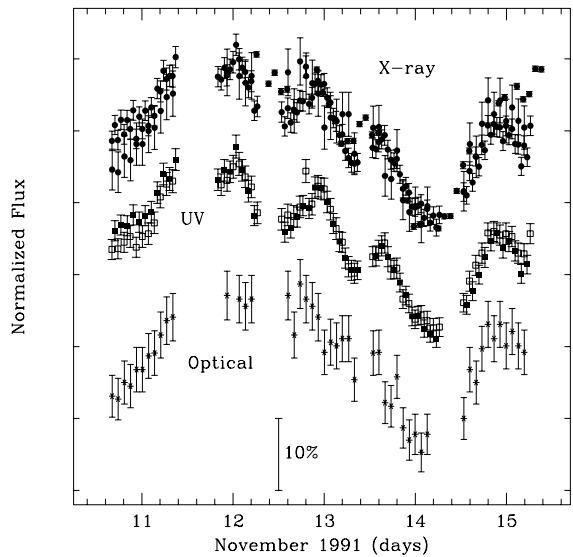,height=3.1in}}
\caption{Normalized X-ray, UV, and optical light curves of PKS 2155--304
from the first in\-ten\-sive mul\-ti\-wave\-length mo\-ni\-tor\-ing campaign, 
in mid-November 1991 (Edelson et al. 1995).
The emission is well-correlated, with comparable amplitude independent
of wavelength, and the X-rays appear to lead the UV by $\sim2$-3 hours.
Even with extensive coverage from the ground, the
best optical data were obtained with the FES monitor on IUE, a simple
star-tracking device, which highlights the value of adding modest optical
devices to high-energy satellites.} 
\label{fig:nov91}
\end{figure}

Results from the November 1991 campaign are shown in Figure~\ref{fig:nov91}.
The X-ray, UV, and optical light curves are well-correlated, 
arguing for a common origin of the optical through X-ray emission,
and the X-rays lead the UV by $\sim2$-3 hours (Edelson et al. 1995). 
The optical/UV emission can not be produced by thermal emission from
a viscous accretion disk because they should arise at very different
radii, implying a large lag between the two.
The amplitude of variation is independent of wavelength, 
a result contrary to what is expected from a synchrotron flare
caused by an increase in energetic electrons
(Celotti, Maraschi, \& Treves 1991).

Several aspects of the 1991 results were intriguing enough to inspire us
to repeat the experiment for a longer period and with more extensive
wavelength coverage. First, the light curves consisted of a series 
of peaks modulated by an overall decline in flux; while not strictly
periodic (Edelson et al. 1995), they suggested possible repetition 
through five cycles.
Second, the UV and X-ray light curves, which tracked very well for most of
the observation, appeared to diverge at the very end of the overlapping period.

\begin{figure}[t]
\centerline{\psfig{figure=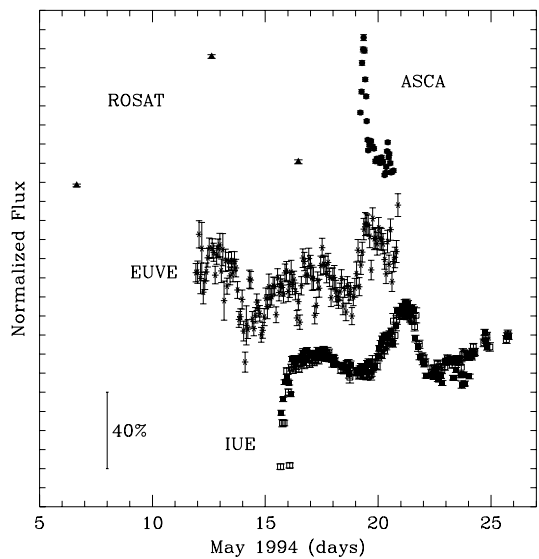,height=3.1in}}
\caption{Normalized X-ray, EUV, and UV light curves of PKS~2155--304
from the second intensive multiwavelength monitoring campaign,
in May 1994 (Urry et al. 1996). 
The ASCA data show a strong flare, echoed one day later
by EUVE and two days later by IUE. The amplitude of the flares decreases 
and the duration increases with increasing wavelength.} 
\label{fig:may94}
\end{figure}

We therefore arranged a second intensive monitoring campaign in May 1994,
with IUE for 10~days continuously (Pian et al. 1996), 
EUVE for 9~days (Marshall et al., in prep.), and ASCA for 2~days
(Kii et al., in prep.),
and with additional Rosat (Urry et al. 1996) and ground-based data
(Pesce et al. 1996).
Figure~\ref{fig:may94} shows the results from this second set of 
multiwavelength observations of PKS~2155--304 (Urry et al. 1996).
A very sharp flare is seen in X-rays, followed by an EUV flare one day later 
and a broader, lower amplitude UV flare two days later. 
The X-ray flux doubles in $\sim0.2$~days and is approximately symmetric. 
The Rosat data, while sparse, show that large X-ray flares were not
unusual but were occurring throughout the week prior to the ASCA 
observations.
The EUV flux increased by $\sim50$\% in less than a day
(the time scale is difficult to estimate given the 
errors and the untimely end of the data train),
and the UV flux rose by $\sim35$\% with a doubling time scale of
$\sim3$~days. The duration of the flare increases from less than a day
in the X-rays to nearly 4 days in the UV.

\begin{figure}
\centerline{\psfig{figure=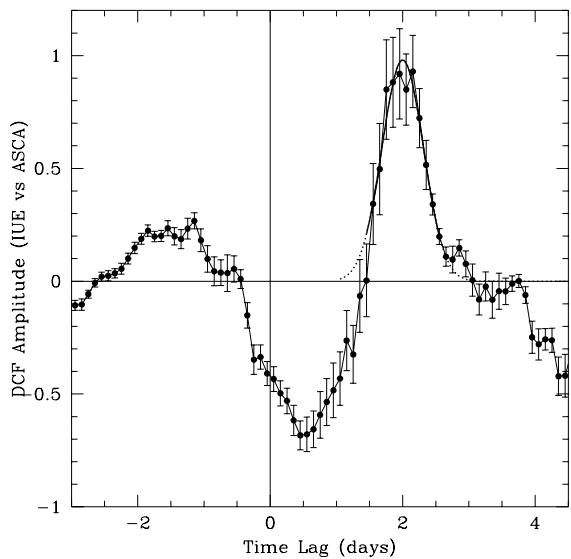,height=3.1in}}
\caption{Cross correlation of May 1994
ASCA and IUE light curves of PKS~2155--304,
showing that the X-ray flux leads the UV by $\sim2$~days
(Urry et al. 1996).}
\label{fig:cc}
\end{figure}

The delays between X-ray, EUV, and UV light curves are easily measured
with cross-correlation functions. Formally, the EUV flux leads the UV
by 1.1 days and the X-ray flux leads the UV by 2.0 days.
The cross-correlation between IUE and ASCA light curves is shown 
in Figure~5; the cross-correlation of X-ray versus EUV light curves 
is not well defined due to the small temporal overlap.
(Fig.~\ref{fig:cc}).

The IUE data also reveal a complex and extremely rapid flare at
the beginning of the observation, with doubling times as fast as 1~hour,
the fastest ever observed at ultraviolet wavelengths and comparable to the
fastest doubling times seen in the X-ray (Pian et al. 1996). 
The event is seen in both LWP
and SWP cameras, and is undersampled by both; it has larger amplitude
in the LWP in part because the LWP integration times are less than half 
the SWP integration times. 
It is also possible to see similar structure in the EUV light curves,
although the EUVE data are relatively noisy, 
$\sim1$-2~days in advance of the IUE event.
\begin{figure}[t]
\centerline{\psfig{figure=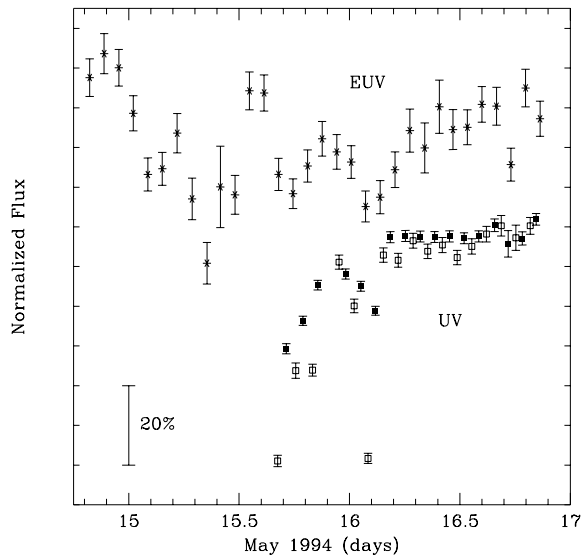,height=3.1in}}
\caption{Expanded view of the beginning of the IUE light curve of 
PKS~2155--304 from May 1994, with the EUVE light curve shifted forward 
by 1.25 days. The extremely rapid UV variability on May 16, undersampled
in both LWP and SWP cameras, is similar to earlier, lower amplitude
variations in the EUVE light curve.}
\label{fig:uv_euv}
\end{figure}
Figure~\ref{fig:uv_euv} shows an expanded view of the EUVE and IUE
light curves, with the EUVE curve shifted forward by 1.25 days.

There are strong differences between the May 1994 and November 1991 
light curves of PKS~2155--304. 
In the second epoch, the amplitude and duration depend
strongly on wavelength and the lags are considerably longer
than in the first. In addition,
there is little of the low-amplitude repetitive variability, at least in
the first half of the IUE observation.
These differences mean either that there are two different
mechanisms operating or that the relevant physical parameters have changed
considerably.
The variability in 1991 was probably not a synchrotron flare,
since the observed time scales were energy independent.
Instead, it is possible that it was caused by microlensing: for a dense
star cluster at the redshift of a known Lyman-alpha absorption system 
(about half way to the BL Lac), and assuming relativistic motion of the
BL Lac jet, the amplitudes and time scales are approximately correct,
and the achromatic nature is automatic as long as the source size is
independent of wavelength.
This is a plausible but not proven explanation.

The 1994 data are much richer in extent and wavelength coverage. These
data are consistent with a synchrotron flare in a jet but 
the clear delay between X-ray and UV flares rules out the homogeneous
case (Urry et al. 1996). The reasoning is as follows.
The simplest causes of a flare in the homogeneous case would be a sudden
uniform increase in the injection of energetic electrons or
the instantaneous and uniform enhancement of the magnetic field,
perhaps via compression of a charged plasma. In both cases, the flux at all
(optically thin) wavelengths would rise simultaneously, with amplitude
increasing with decreasing wavelength, while the duration
of the flare (if due to energy losses only)
would go as $\lambda^{1/2}$. That is, the long-wavelength emission would
last longer than the short-wavelength emission but
the (instantaneous) peaks in the light curves would be simultaneous.
The delay in the flare onset and the longer rise time for the UV
flare compared to the X-ray flare clearly rule out the simplest
homogeneous case, unless {\it ad hoc} dependences on time and energy of
the injection rate are postulated.

Thus, some energy stratification of the synchrotron-emitting plasma
is required, either behind a shock or in the jet structure itself.
The X-ray decay is faster than the UV but perhaps by less than a factor 
$(\lambda_{UV}/\lambda_{X})^{1/2}\sim20$, suggesting the
decay is dominated by geometry rather than synchrotron losses.
The flare duration increases with wavelength, possibly indicating that 
the size of the emitting region is increasing with wavelength.
The observed lags are also comparable to the flare durations.
Both these results are as
expected for a shock propogating outward in an inhomogeneous jet,
successively passing from X-ray- to EUV- to UV-emitting regions. 

\section{Future Multiwavelength Monitoring}

As this review has shown, the best available data are still insufficient
for determining jet structure but enough to indicate the kind of data needed.
We should repeat the kind of multiwavelength monitoring done for 
PKS~2155--304 for many more blazars, including LBL, which will require 
a new ultraviolet capability. We also need to study further the correlation
of gamma-ray and optical/IR variability. In all cases,
long and intensive time sampling is critical: 
two light curves that are well correlated
could, if sampled at few points or for less time than the characteristic lag,
appear uncorrelated. Thus the single epoch approach is no longer
valuable for adding information. The requirement for intensive monitoring and
for larger samples of objects points to the need for a multiwavelength 
platform with modest, very simple, very inexpensive optical/UV telescopes
paired with X-ray and gamma-ray detectors.

\acknowledgments

I am grateful to my many collaborators for their contributions to our
multiwavelength monitoring of blazars, particularly those with whom
I have worked most closely in recent years, Laura Maraschi, Joe Pesce, 
Elena Pian, Rita Sambruna, Aldo Treves, and Ann Wehrle, and those who were 
PIs for the second campaign on PKS~2155--304, Tsuneo Kii, Herman Marshall, 
and Greg Madejski.

\begin{question}{Al Marscher}
In the 1994 PKS~2155--304 flare, you say that the fast rise is from the 
time scale for particle acceleration but that the slower decay is geometric.
Light travel effects should limit the shorter time scale, not the longer. 
Also I disagree that the longer time scales and lags at lower frequencies 
require gradients in the underlying jet to be the root cause: shocks 
should be frequency-stratified and give the behavior you observe 
(Marscher, Gear \& Travis 1992, in {\it Variability of Blazars}).
\end{question}
\begin{answer}{Meg Urry}
You are right about the geometric limit for the fastest time scale.
We agree that the jet needs to be inhomogeneous. Whether the inhomogeneity
arises from the natural stratification due to energy-dependent losses and
diffusion from a shock, or from some other means, is not known at this
point.
\end{answer}
\begin{question}{Hugh Aller}
I am worried about involving different physical processes to explain
two different flare events (1991 versus 1994) in PKS~2155--304.
Isn't a single process which permits both events to be explained
ultimately simpler?
\end{question}
\begin{answer}{Meg Urry}
I can see how it would seem that way but frankly, the character of the
variations in 1991 and in 1994 is so different that the required change
in parameters of the single process you might prefer is more or less
equivalent to a distinct process. If, for example, we decide that the
1994 flares are the signature of a shock propagating through an 
inhomogeneous jet, with the expected dependence of amplitude on wavelength
and delays of $\sim1$~day from X-rays to EUV to UV-emitting regions,
then how can this model explain the 1991 variability, with its achromaticity
and negligible lags? In a sense these are opposite behaviors, which
naturally suggest unrelated causes.
\end{answer}

\end{document}